\documentclass[conference]{IEEEtran}
\usepackage{xr}
\usepackage{balance}
\usepackage[noadjust]{cite}
\usepackage{subcaption}
\usepackage{stackengine}
\usepackage[pdftex]{color,graphicx}
\usepackage[bookmarks=false]{hyperref}
\usepackage{float}
\usepackage{graphicx,amsmath,url}
\usepackage{acronym}
\usepackage{flushend}
\usepackage{booktabs}
\ifCLASSINFOpdf
\else
\fi
\acrodef{CMOS}[CMOS]{complementary metal-oxide-semiconductor}
\acrodef{NEMS}[NEMS]{nano electro-mechanical switches}
\acrodef{NEM}[NEM]{nano electro-mechanical}
\begin{document}
%
\title{Energy-efficient Hybrid CMOS-NEMS LIF Neuron Circuit in 28\,nm CMOS Process}

\author{\IEEEauthorblockN{Saber Moradi}
\IEEEauthorblockA{Computer Systems Laboratory\\
Yale University,
New Haven, CT 06520\\
saber.moradi@yale.edu}
\and
\IEEEauthorblockN{Sunil A. Bhave}
\IEEEauthorblockA{School of Electrical and Computer Engineering\\
Purdue University, Indiana 47907-2035\\
bhave@purdue.edu}
\and
\IEEEauthorblockN{Rajit Manohar}
\IEEEauthorblockA{Computer Systems Laboratory\\
Yale University,
New Haven, CT 06520\\
rajit.manohar@yale.edu}}

\maketitle

\begin{abstract}
  
Designing analog sub-threshold neuromorphic circuits in deep sub-micron
technologies~e.g.~28\,nm can be a daunting task due to the problem of excessive leakage current. 
We propose novel energy-efficient hybrid CMOS-\ac{NEMS} Leaky Integrate and Fire~(LIF) neuron and synapse circuits and investigate the impact of NEM switches on the leakage power and overall energy consumption. 
We analyze the performance of biologically-inspired neuron circuit in terms of leakage power consumption and 
present new energy-efficient neural circuits that operate with biologically plausible firing rates. Our results show the proposed CMOS-NEMS neuron circuit is, on average, 35\% more energy-efficient than its CMOS counterpart with same complexity in 28\,nm process. Moreover, we discuss how NEM switches can be utilized to further improve the scalability of mixed-signal neuromorphic circuits. 

\end{abstract}

\IEEEpeerreviewmaketitle

\section{Introduction}
\label{sec:intro}
Machine learning techniques based on neural networks~\cite{machine_learning_14, machine_learning_17} have recently exhibited state of the art algorithmic performance in many cognitive tasks, such as perception and recognition. These models are usually composed of massively parallel network of computational modules and require a vast amount of memory and computation. Due to these requirements, running neural network models on von Neumann based machines can only exacerbate the memory bottleneck of conventional computer architectures, limiting their overall performance. Neuromorphic computing~\cite{mead90}, inspired by biological neural systems, offers radically different architectures and models for performing brain-like computation. Highly parallel platforms with co-located memory and computation are central to the advancements of neuromorphic applications. Such platforms allow us to explore various networks configurations and parameters. This has been a key motivation for research groups in academia and industry to aim to build large-scale (multi-million neurons) networks. \\While there have been successful demonstrations of sub-100\,nm neuromorphic systems in the digital domain~\cite{merolla2014million}, mixed signal designs face more severe challenges in deep sub-micron technologies~\cite{analog_scaling}. This is mainly due to the leakage current of field effect transistors becomes a significant portion of transistor's on-current in more advanced \ac{CMOS} processes, leading to higher power consumption. In neuromprhic digital designs, the impact of leakage power is reduced by multiplexing the circuit blocks and therefore avoiding excessive silicon area. However, the problem is aggravated in highly-parallel mixed-signal neuromorphic circuits that operate in biological time scales. In these circuits, the leakage power becomes the dominant factor as the active power is very low. 
While most neuromorphic systems have employed CMOS technology for
implementing neural principles, there have been significant
research efforts~\cite{hu2014memristor, Serrano-Gotarredona_etal13, sung_etal10} to use emergent nano-devices
e.g. memristive technologies in neuromorphic systems. In particular,
memristive devices theoretically offer a promising way to implement
synapses by providing very compact analog memory elements. The idea
within this context is to use a grid of synaptic memristive devices on
the top of CMOS implemented neuronal arrays. Although, the efficiency
of this combination is demonstrated for single device or a small
array~\cite{hu2014memristor}, scalabiliy of this approach is still being explored in
academia primarily due to the reliability and immaturity of the
technology. In this paper, we aim to use nano-devices in a novel
way---instead of using these devices for computation we utilize a
certain type of NEMS device to address some of the major issues
(e.g. excessive leakage power) that impact scaling CMOS to very
small feature sizes and scaling up the system using a massively
parallel architecture.\\The Leaky Integrate and Fire~(LIF) neuron is a widely used model in neuromorphic community~\cite{gineuron, accl_neuron, mead89, LIF_Arthur04, Aamir2016, moradi2014, qiao2015reconfigurable}. The LIF neuron circuit operating with biologically-realistic firing rates usually requires higher~\cite{gineuron, Aamir2016} energy per spike in comparison to its accelerated-time counterparts~\cite{accl_neuron}.
In this paper, we present an energy-efficient hybrid COMS-NEMS circuits for the LIF neuron and an excitatory synapse model~\cite{moradi2014, cxquad}. Through our analysis of these circuits, we identify the major source of energy consumption and investigate the impact of NEM switches on the leakage power and overall energy consumption of mixed-signal neuromorphic designs. These switches can be integrated with \ac{CMOS}, and resemble mechanical relays rather than field-effect devices. \ac{NEMS} can provide near-zero idle power consumption~\cite{nems_scaling}, thereby potentially enabling higher degrees of parallelism in a neuromorphic computing systems. The main drawbacks of NEM devices, namely limited switching lifetime and slow switching cycle, do not cause major issues in neuromorphic systems with biologically plausible time-constants. \\
The ideas presented in this paper can be applied to other \ac{CMOS} neuron and synapse circuits, however, in order to make concrete comparisons we consider the LIF neuron circuits~\cite{gineuron, qiao2015reconfigurable, moradi2014} as the baseline to illustrate our approach. We show simulation results for the neuron circuit at different operating points and explain the performance improvements that are made possible by the new design.  
The rest of this paper is organized as the following. Section II provides an overview of emerging \ac{NEM} switches. In Section III, we present hybrid CMOS-NEMS circuit diagram for the LIF neuron model, and detail description of the circuit. The performance of the proposed neuromorphic circuit are presented in Section IV. The conclusion and the future works are presented in Section V.

\section{Emerging Nano Electro-mechanical switches~(NEMS)}
\label{sec:nems}
\begin{figure}
\centering
\includegraphics[scale=0.48]{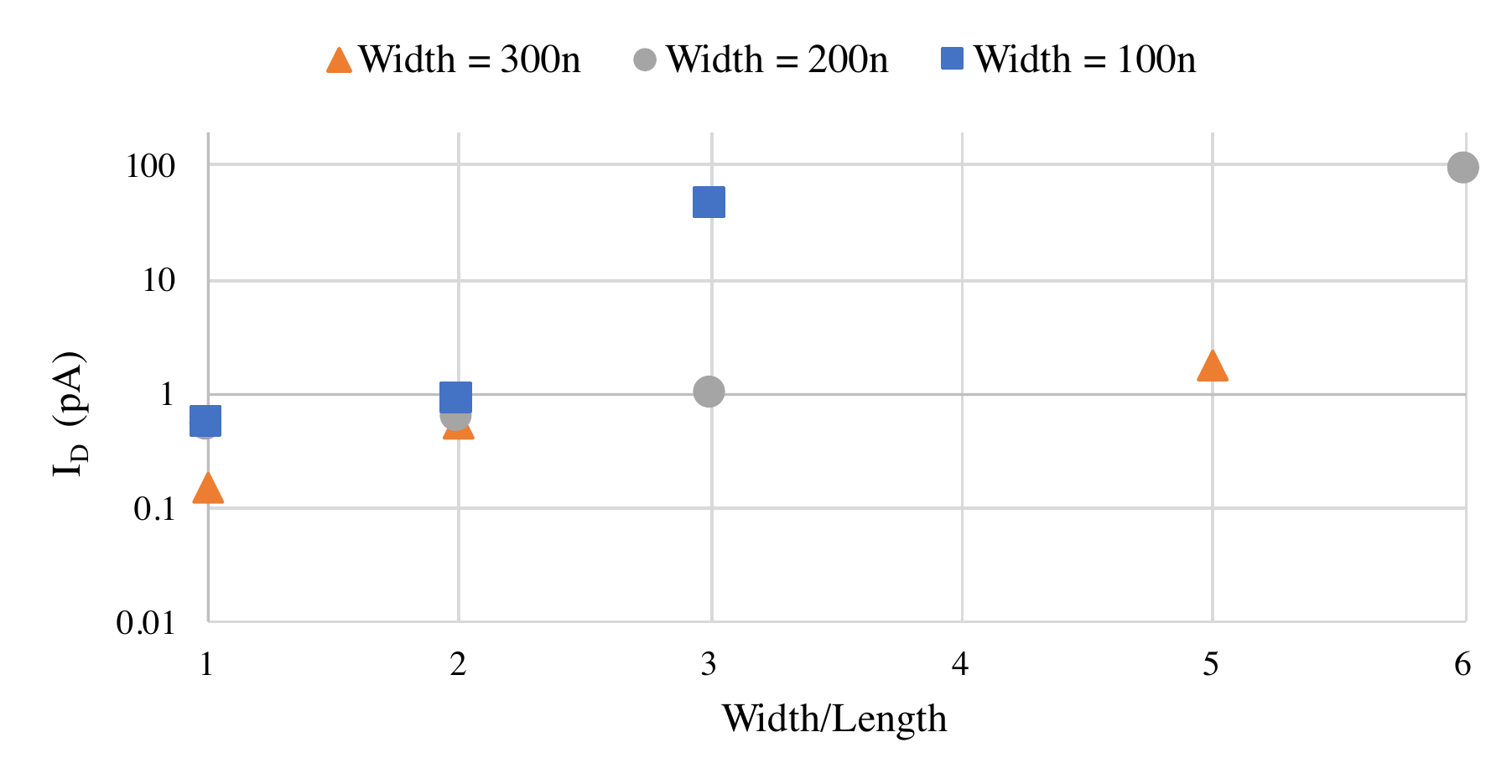}
\caption{Drain current~($I_{D}$) NMOS transistors with different sizes in 28\,nm process while $V_{gs}=0$, $V_{DS}=0.5V$.}
\label{fig:offcurrent}
\end{figure}

\begin{figure}
\centering
\includegraphics[width=1.7in]{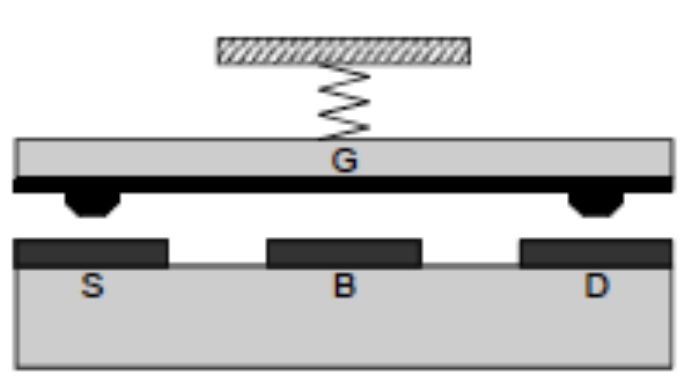}
\caption{NEM relay showing a normally-open topology.}
\label{fig:nemsrelay}
\end{figure}

\begin{figure}
\centering
\includegraphics[width=2.6in]{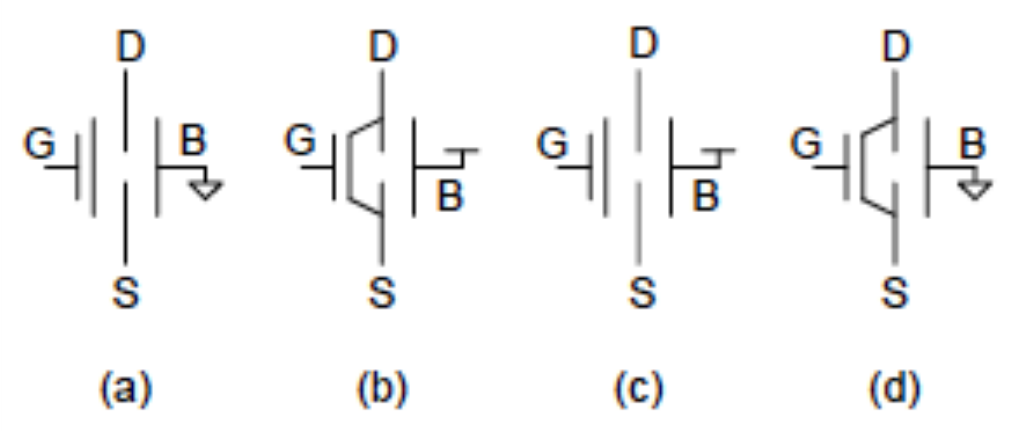}
\caption{NEMS relays used as switches~\cite{async_nems}. (a) ``NMOS'' with
  normally-open relay; (b) ``NMOS'' with normally closed relay; (c)
  ``PMOS'' with normally-open relay; (d) ``PMOS'' with normally-closed
  relay.}
\label{fig:nemsfet}
\end{figure}

The off current of transistors causes energy dissipation when they are not active. The problem gets worse in 
advanced CMOS processes e.g. 28nm where the off current is no longer neglectable in comparison to on current of the transistors, leading to excessive energy consumption. This issue is even more critical in mixed-signal neuromorphic designs where the analog circuits, typically operate with pico-ampere range currents, are integrated with digital switches. As shown in Fig.~\ref{fig:offcurrent}, the off current of NMOS transistors in 28\,nm can be as high as pico amperes even for long transistors~(e.g. Length=300\,nm). High leakage current not only increases the energy consumption but also narrows the range of currents in which the neuromorphic sub-threshold circuits can reliably operate.\\    
NEM switches have been recently been
investigated as a potential alternative to CMOS. While NEMS
are much slower than CMOS, their near-zero off current means that they
enable significantly higher degrees of parallelism.
Prior work~\cite{async_nems} studied four-terminal NEM relays and their impact on
CMOS logic. The same NEMS model is used in our analysis. A brief description of the NEM switches based on this model is presented in the following. Figure~\ref{fig:nemsrelay} shows a normally-open
NEM relay structure, which is a four-terminal switch consisting of a
gate (G), drain (D), source (S), and body (B) terminals. When there is
a large enough voltage difference (called the {\it pull-in voltage\/})
across the gate and body ($V_{gb}$), the electrostatic force generated is large
enough to cause the suspended structure to be pulled toward the body
thereby connecting the source and drain terminals. When $V_{gb}$ is
smaller than the pull-in voltage, the gate pulls away from the body
due to the spring's restoring force. A similar principle is used for
a normally closed NEM relay. In this case, the default switch state
is on, and when $V_{gb}$ exceeds the threshold voltage, the source and
drain terminals are disconnected. 
Normally-open and normally-closed NEM relays can be used to build
replacements for NMOS and PMOS switches. Fig.~\ref{fig:nemsfet}
shows four configurations that implement the two types of transistors
with the two different relay types. A challenge with the ideal
scenario depicted here is that manufacturing uncertainty can require
{\it different\/} body voltages for different NEMS devices. Compared to digital CMOS switches, 
the switching delay of NEMS is high, and the number of time a device switches is limited before it fails. 
A benefit of the NEMS architecture is that both the ``NMOS'' and
``PMOS'' replacement structures can conduct high and low voltages
equally well. Hence, it is easy to build non-inverting logic gates
with NEM switches. Also, the body can be exposed as a terminal and used
for useful computation making gates like XOR and XNOR particularly
inexpensive in device count compared to their fully restoring CMOS
counterparts.

\section{Hybrid CMOS-NEMS Neuromorphic Circuits}
\label{sec:circuits}
\begin{figure}
\centering
\includegraphics[scale=0.28]{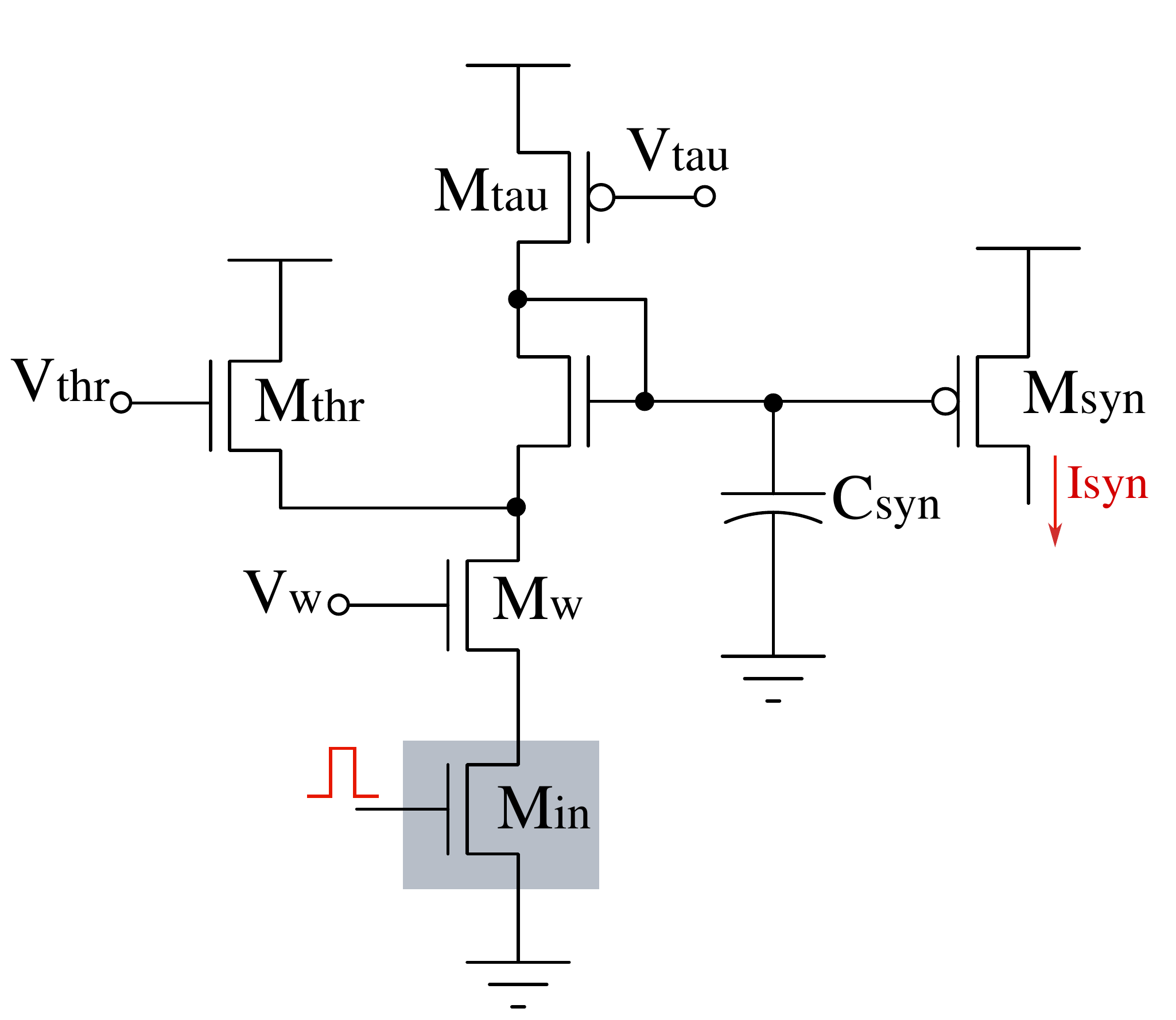}
\caption{CMOS-NEMS Diff-pair Integrator~(DPI) synapse circuit~\cite{moradi2014}. $M_{in}$ is replaced with a NEM switch to turn off input transistors when there is no input activity.}
\label{fig:nemynz}
\end{figure}

In this section, we present new CMOS-NEMS neuron and synapse circuits and show how NEM switches can be utilized to reduce energy consumption of such circuits. In particular, we present hybrid CMOS-NEMS neuromorphic circuits for Differential-pair Integrator~(DPI) synapse circuit and the LIF neuron model~\cite{moradi2014, cxquad}, although same concept can be applied to other neurons and synapse models.\\
Figure~\ref{fig:nemynz} illustrates a current-mode circuit for implementing the dynamics of excitatory synapses. $M_{in}$ transistor acts as enable signal for the synapse circuit. In the absence of an input, $M_{in}$ is turned off and capacitor $C_{syn}$ is charged to $ Vdd $ through $M_{tau}$. This condition turns $M_{syn}$ off and therefore the synaptic output current $I_{syn}$ to near zero. Upon receiving a new event, $C_{syn}$ starts discharging through $M_{in}$ and $M_{w}$, proportional to the $M_{w}$ current(set by weight bias $V_{w}$) and the duration of input pulse. After the input pulse ends, $M_{tau}$ begins to charge $C_{syn}$ again with a current amplitude set by $V_{tau}$. Given that the input pulse is in normally order of sub-microseconds and the synaptic time constant can be in order of hundreds of milliseconds, unwanted leakage current of $M_{in}$ increases power consumption and in some cases can cause misbehavior of the circuit. We have replaced $M_{in}$ with a normally open ``NMOS'' type \ac{NEM} relay. 
\begin{figure}
\centering
\includegraphics[scale=0.96]{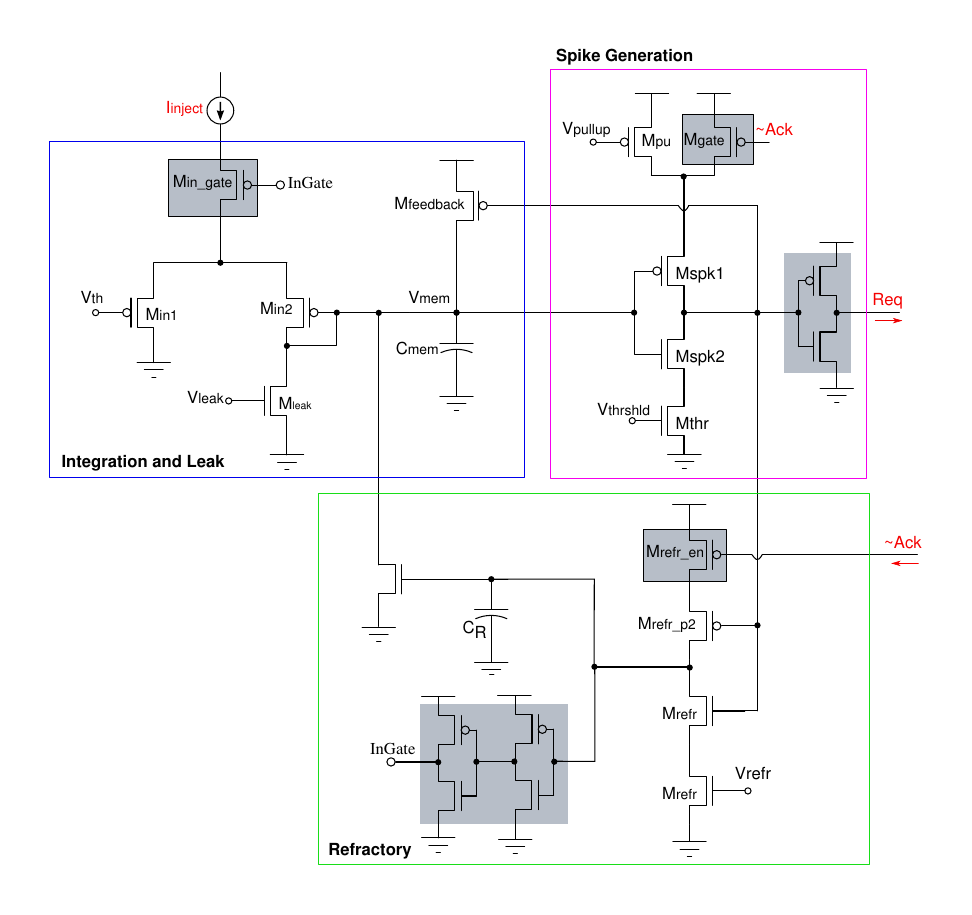}
\caption{The proposed LIF neuron circuit. The highlighted transistors act as digital switches and while all other transistors operate in analog sub-threshold regime. The digital switches are replaced by NEMS. These switches are used for power-gating parts of the circuit that are inactive. Power gating ``comparison branch'' using $M_{gate}$ significantly reduces the energy usage by limiting the current during the integration phase. The size of capacitors are $C_{mem}$ is 500\,fF and $C_{R}$ is 150\,fF.}
\label{fig:circuits}
\end{figure}
The NEM switches are combined with CMOS circuits to design an energy-efficient CMOS-NEMS LIF neuron. In the following, detailed description of the circuit and the way NEMS switches are utilized to improve energy efficiency are presented. Fig.~\ref{fig:circuits} shows the schematic of the LIF neuron circuit. The highlighted transistor are used as digital switches while all other transistors are biased in analog sub-threshold regime. All highlighted transistors are replaced with NEM switches in hybrid CMOS-NEMS circuit. The circuit involves three parts: ``integration and leak'', ``spike generation'' and ``refractory''. The ``comparison branch'' is a part of ``spike generation'' and includes $M_{Spk1-2}$, $M_{thr}$ and $M_{pu}$ transistors.
In the first phase, the membrane capacitor $C_{mem}$ is charged by the input current $I_{inject}$ while $M_{leak}$ discharges the membrane. As long as the injection current is stronger than $I_{leak}$, the membrane voltage $V_{mem}$ starts increasing until it gets close to the neuron threshold voltage--set by $V_{thrshld}$. At this point $M_{Spk2}$ slowly begins to lower down the output of ``comparison branch''; this transition is made quicker by a feedback mechanism implemented with $M_{feedback}$. Once the output of the ``comparison branch'' flips, the ``spike generation'' circuit issues a spike indicating a new event being generated. Simultaneously to generating an event, the refractory capacitor $C_{R}$ is charged through $M_{refr\_en}$-$M_{refr\_p2}$ transistors and consequently resets $ V_{mem} $ to the resting potential(i.e. zero here). Once the event is acknowledged and the neuron is reset by the refractory circuit, $M_{gate}$ is turned on and sets the output of ``comparison branch'' to high again.\\
The comparison branch in the ``spike generation'' part is a major source of energy consumption in the neuron circuit as for a long period of time (in orders of several ms), $V_{mem}$ voltage value is around the threshold voltage of $M_{Spk1}$and $M_{Spk2}$. This causes a static current to flow through ``comparison branch'' and therefore an increase in overall energy consumption.  
To reduce the energy consumption, the spike generation circuit are power-gated using a NEM switch $M_{gate}$ driven by the output handshaking signal~(ACK). Therefore the P network of the comparison circuit is disabled, limiting the static current going through comparison circuit during the integration phase. A weak pull-up transistor $M_{pu}$ is used to bias the circuit. 
 Additionally, we introduced a \ac{NEMS} switch at the input to block the input current current while a new event being generated and the refractory circuit is active. The proposed hybrid CMOS-\ac{NEMS} neuromorphic circuits is simulated under different conditions and the results are presented in the following section.

\externaldocument[I-]{circuit}
\section{Simulation Results and Analysis}
\begin{figure}[t!]
\centering
\includegraphics[scale=0.48]{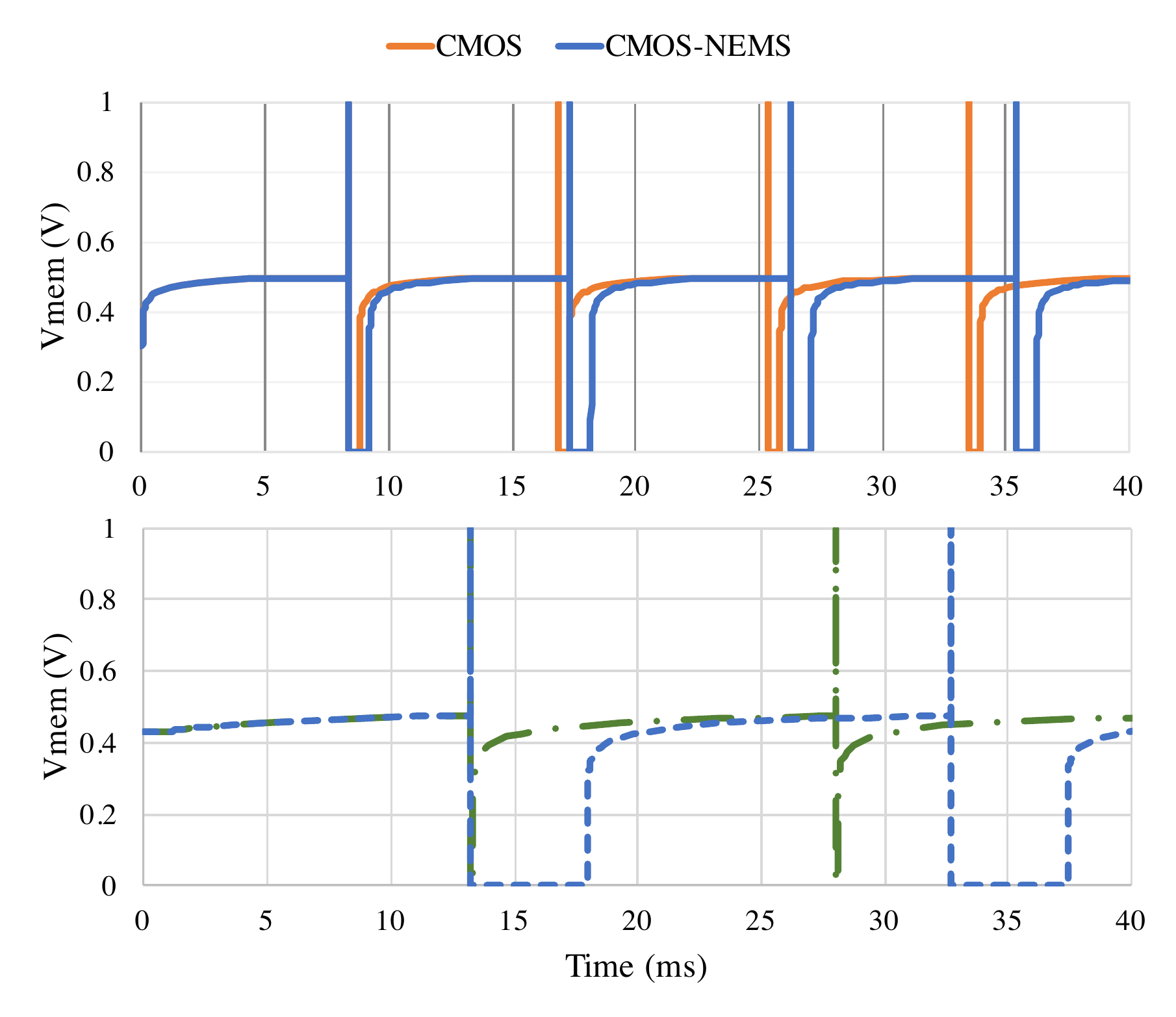} 
\caption{Top: neuron's membrane voltage $V_{mem}$ of the CMOS and CMOS-NEMS circuits. The injection current is set to 0.25\,nA and the $V_{refr}$ bias is adjusted for the refractory period of 0.53\,ms. Bottom: membrane voltage of CMOS-NEMS neuron circuit with two different refractory periods. }
\label{fig:membrane}
\end{figure}
\begin{figure}[b]
\centering
\includegraphics[scale=0.44]{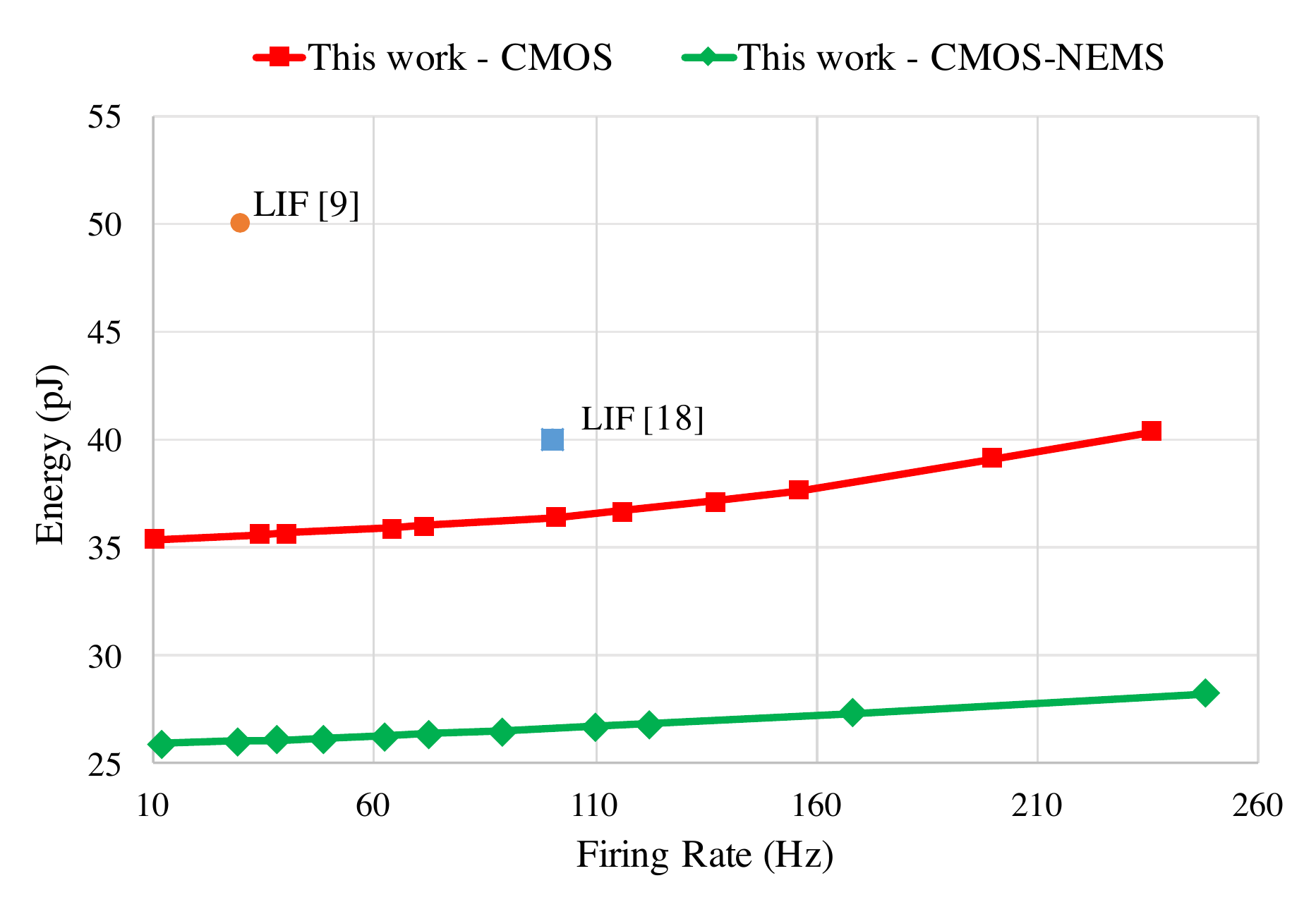} 
\caption{Energy comparison: the proposed CMOS-NEMS neuron circuit vs the CMOS and existing implementations~\cite{qiao2015reconfigurable, comp_2} in 28\,nm. Thanks to the NEM switches used for power-gating, the proposed CMOS-NEMS circuit on average requires 35\% less energy to operate with the same firing rates.}
\label{fig:nemsenergy}
\end{figure}
\begin{figure}
\centering
\includegraphics[scale=0.47]{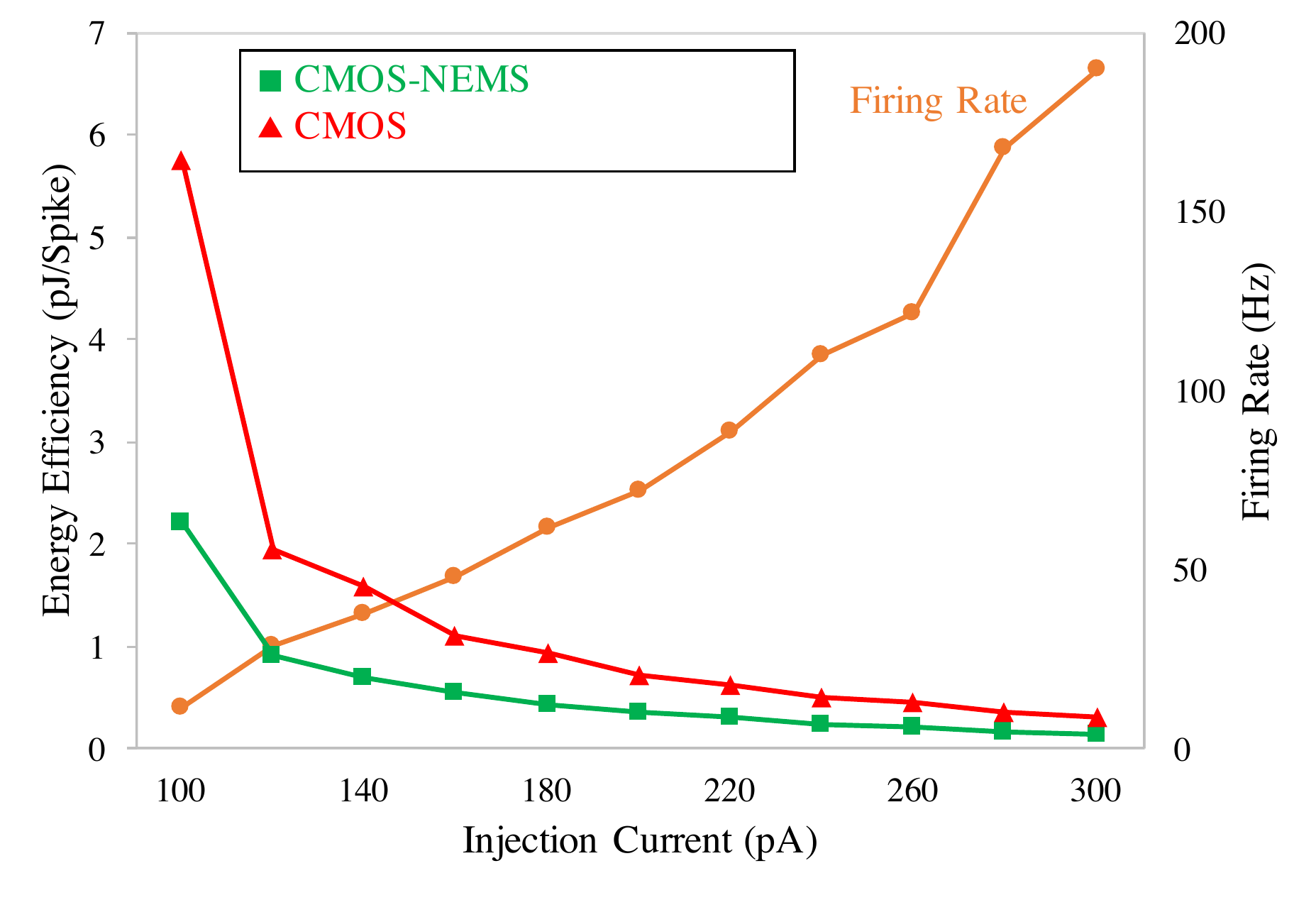}
\caption{Energy Efficiency~(pJ/spike) vs injection current~(pA). The hybrid CMOS-NEMS circuit has better energy-efficiency in low firing rates, making it more suitable to use in large-scale neuromorphic systems operating with biologically-realistic time-constants.}
\label{fig:energyeff}
\end{figure}
We have run simulations for the proposed \ac{CMOS}-\ac{NEMS} neuron circuits and compared its performance with its CMOS counterpart and that of~\cite{qiao2015reconfigurable, comp_2}. The \ac{NEM} switches in these simulations are based on Verilog-A model from~\cite{async_nems, nems_chen}. In the following text, the CMOS circuit refers to Fig.~\ref{fig:circuits} circuit except for the NEM switches~($M_{in\_gate}$, $M_{gate}$ and $M_{refr\_en}$), which is also equivalent to the LIF neuron circuit in~\cite{qiao2015reconfigurable} without the adaption feature.  
Fig.~\ref{fig:membrane}~(top) shows a trace of the membrane voltage $V_{mem}$ of the CMOS and hybrid CMOS-NEMS circuits where injection current is set to 250\,pA in pure CMOS and 235\,pA in the hybrid circuit. Additionally $V_{refr}$ in both circuits is adjusted to get similar refractory time~($\sim$0.53\,ms) after each spike. Both circuits operate with the same firing rate and refractory time. Figure~\ref{fig:membrane}~(bottom) illustrates the membrane voltage of the CMOS-NEMS circuit for very short~(ns) and long~(several ms) refractory periods.\\
The energy usage of both circuits under different firing rates~(10-250\,Hz) is illustrated in Fig.\ref{fig:nemsenergy}. The results indicate that CMOS-NEMS neuron circuit is on average 35\% more energy efficient than the CMOS circuit when operating with biologically plausible firing rates. 
This improvement is made possible largely due to the NEM switches, used to power-gate parts of the circuits which are inactive~($M_{in\_gate}$, $M_{gate}$ and $M_{refr\_en}$). In particular $M_{gate}$ switch plays a major role in the energy reduction as the ``comparison branch'' takes a big portion of the overall energy. This is because for a long period of time~(in orders of several ms) $V_{mem}$ is near the threshold voltage, leading to flow of static current through the ``comparison branch''. The $M_{gate}$ switch turns off the P network of comparison branch during the integration phase and therefore limits the static current of the branch.
The circuit's energy efficiency versus the input injection current is illustrated in Fig.~\ref{fig:energyeff}. The results indicates that CMOS-NEMS neuron circuit is more energy efficient for biologically plausible spike activities. Such energy-efficiency combined with recent advances to reduce the NEMS footprint, makes hybrid CMOS-NEMS a viable solution to design energy-efficient large-scale biologically plausible neuromorphic systems. Finally, we have compared the proposed circuit in this paper with existing LIF neuron implementations~\cite{qiao2015reconfigurable, comp_2} in terms of silicon area and energy requirements~(see Table~\ref{tab:table1}). In this comparison, the size of a NEM switch is considered $0.1um^{2}$~\cite{nems_scaling, nems_scaling_3}. The silicon area of the LIF neuron is dominated by the the capacitors. While new techniques~\cite{nems_scaling, nems_scaling_2} have been proposed to scale down the footprint of NEM devices, it should be noted that to achieve low-mismatch and low-leakage behavior in subthreshold regime, the size of CMOS transistors need to be much larger than the minimum feature size in the advanced processes.\\

\begin{table}[ht]
  \centering
  \caption{LIF neuron circuits: performance comparison.}
  \label{tab:table1}
  \begin{tabular}{ccccc}
    \toprule
     & Process & Area~($ um^{2}$ ) & Firing Rate~(Hz) & pJ/spike\\
    \midrule
    LIF~\cite{qiao2015reconfigurable} & 28\,nm & 70 & 30  & 1.7\\
    LIF~\cite{comp_2} & 90\,nm& 442  & 100 & 0.4\\
    This work--CMOS  & 28\,nm & $\sim$70  & 30/100 & 1.17/0.36\\
    This work--Hybrid  & 28\,nm & $\sim$80  & 30/100 & 0.84/0.25\\
    \bottomrule
  \end{tabular}
\end{table}
The neuron firing rates for biologically plausible neurons is typically 10\,Hz~\cite{gineuron}. This means that the NEM switches utilized in a neuron will have to cycle 10 times per second. This is much slower than the requirements for high-speed digital logic. NEM switches with lifetimes of $ 10^{10} $ cycles~\cite{nems_scaling_2}--\cite{nems_scaling_3} have been demonstrated, and this translates to a chip lifetime of roughly 30 years for neuromorphic applications. Hence, using NEMS for neuromorphic systems is a good match in terms of the characteristics of NEM switches and the requirements for neuromorphic circuits.
\section{Conclusion}
We have presented a new hybrid CMOS-NEMS LIF neuron circuit that is $\sim$35\% more energy-efficient than existing CMOS designs with same circuit complexity. We have studied and analyzed the performance of the LIF neuron CMOS circuits in terms of energy consumption and proposed a new circuit to further reduce the energy consumption. We have shown how NEMS switches can be combined with the neuromorphic CMOS circuits. Our results indicates that utilizing the \ac{NEM} relays instead of digital CMOS switches significantly reduces the energy usage of LIF neuron circuits operating with biologically realistic time-constants. Moreover, the \ac{NEMS} devices can used for implementing local inter-neuron connectivity in large-scale neuromorphic computing systems. Combined with the recent advancement in improving switching life-time of NEMS switches and their footprint, hybrid CMOS-NEMS approach has great potential to become a viable technology for developing energy-efficient neuromorphic architectures.

\end{document}